# MVP: Detection of motif-making and -breaking mutations

Afif Elghraoui[1,2] and Faramarz Valafar[1]*

1. Laboratory for Pathogenesis of Clinical Drug Resistance and Persistence, San Diego State University
2. Department of Electrical Engineering, University of California, San Diego

* Corresponding Author: faramarz@sdsu.edu

# Abstract

**Background:**
DNA, RNA, and protein sequence motifs can be recognition sites for biological functions such as regulation, DNA base modification, and molecular binding in general. The gain and loss of such motifs can carry important consequences. When comparing sequences, the analysis of individual variants does not impart an understanding of the impact on these sites. Rather, only when these variants considered together with their neighbors and the original sequence context does this become possible.

**Results:**
The motif-variant probe (*mvp*) makes this consideration, counting instances of user-specified sequence motifs before and after mutation and reports those that result in motif gain or loss. *mvp* can perform a similar analysis for proteins with amino acid variant data. The software is freely available at https://lpcdrp.gitlab.io/mvp and also installable with the conda package manager.

**Conclusions:**
The ability to easily search for variants affecting any motif, together with the simultaneous consideration of neighboring variants makes *mvp* a versatile tool to aid in a less-frequented dimension of comparative genomics.

# Background

Molecular binding sites on DNA and proteins are often represented by sequence patterns or motifs[1, 2]. Any variation in these motifs can disrupt binding interactions, making the detection of such variation an important goal. Rather than simply counting the differences in motif occurrence between two sequences, knowing the responsible mutations facilitates their use as potential phenotypic markers. FunSeq2[3] has similar functionality to *mvp*, but is oriented towards cancer genomics with its consideration of pre-defined transcription factor position weight matrices (PWMs).

*mvp* requires a motif to search for and, while tools such as MEME[4] traditionally satisfy this requirement, third-generation sequencing has enabled more widespread collection of epigenomic data, including the motif predictions based on where the sequencer detects base modifications, making a need for such analyses more frequent. *mvp* has already been applied in a comparison of *Mycobacterium tuberculosis* reference strains[5], as well as in a study of methylation motif conservation in *M. tuberculosis* clinical isolates[6].

# Implementation

*mvp* takes as input genomic variants in the variant call format (VCF)[7], the reference sequence (fasta format) with respect to which the variants are called, and a set of motifs to investigate. Internally to the program, motifs are represented as regular expressions to be queried against a biological sequence.

Based on the length *L* of the motif of interest, an appropriate minimal sequence context size to search for is determined by considering the cases of two possible extremes. In one extreme, the mutant base represents the last base in the motif, so, to see this motif, we need to examine the preceding *L*-1 bases. In the second extreme, the mutant base represents the first base in the motif, so we also examine the subsequent *L*-1 bases to enable detection of this case. Thus, for each variant, motif occurrences are scanned in the sequence segment centered on the variant and extending *L*-1 bases in each direction. If another variant occurs within this range, the range is extended a further *L*-1 bases and the variants are considered simultaneously, as they may both be affecting the same motif instance. Instances of the query motif are counted in this and compared to the corresponding reference segment. If the counts differ, *mvp* has found a motif-variant and reports it in the output together with the motif counts in the reference and in the sample.

## Conclusions

Pinpointing the gain and loss of specific motif instances in DNA, RNA, and protein can provide insight into mechanisms of adaptation. By doing so through highlighting the responsible mutations, *mvp* enables the targeted study of subtle evolutionary events.

## Availability and requirements

Project name: mvp

Project home page: https://lpcdrp.gitlab.io/mvp

Operating system(s): Platform independent

Programming language: Python

Other requirements: pysam

License: GNU GPL

Any restrictions to use by non-academics: N/A

## List of abbreviations

PWM: position weight matrix

VCF: variant call format

## Declarations

### Ethics approval and consent to participate

Not applicable

### Consent for publication

Not applicable

### Availability of data and materials

Not applicable

## Competing interests
The authors declare that they have no competing interests.

## Funding
This work was funded by a grant from National Institute of Allergy and Infectious Diseases (NIAID Grant No. R01AI105185).

## Authors' contributions
AE conceived the project, wrote, and tested the program. AE and FV prepared the manuscript.

## Acknowledgements
We thank Calvin C. Kim for code contribution and testing.